\documentclass[prl,aps,twocolumn,showpacs,amsmath,amssymb]{revtex4}
\usepackage{graphicx}
\usepackage{dcolumn}

\begin{document}

\title{Evolution of SU(4) Transport Regimes in Carbon Nanotube Quantum Dots}
 
\author{A. Makarovski,$^{1}$ J. Liu,$^{2}$ and G. Finkelstein$^{1}$}

\affiliation{ Departments of  $^1$Physics and $^2$Chemistry, Duke University, Durham, NC 27708}

\begin{abstract}
We study the evolution of conductance regimes in carbon nanotubes with doubly degenerate orbitals (``shells'') by controlling the contact transparency within the same sample. For sufficiently open contacts, Kondo behavior is observed for 1, 2, and 3 electrons in the topmost shell. As the contacts are opened more, the sample enters the ``Mixed Valence'' regime, where different charge states are strongly hybridized by electron tunneling. Here, the conductance as a function of gate voltage shows pronounced modulations with a period of four electrons, and all single-electron features are washed away at low temperature. We successfully describe this behavior by a simple formula with no fitting parameters. Finally, we find a surprisingly small energy scale that controls the temperature evolution of conductance and the tunneling density of states in the Mixed Valence regime.

\end{abstract}


\maketitle

At low temperatures, carbon nanotubes demonstrate a rich variety of transport regimes \cite{CNTreview}. In samples with low contact transparency, a pronounced Coulomb blockade \cite{QDreview} is observed, where single-electron conductance peaks are separated by broad ``valleys'' of vanishing conductance \cite{Tans-Bockrath97}. A more complex behavior is revealed in samples with larger contact transparency, where the conductance is partially blocked, but the signal in the valleys with a non-zero electron spin increases at low temperatures \cite{Nygard2000}, manifesting the Kondo effect. This many-body effect \cite{Hewson}, has recently been observed in a variety of nanoscale systems, including semiconductor quantum dots \cite{Kondo1998}, molecules, carbon nanotubes, and magnetic addatoms on metallic surfaces (see Ref. \cite{revival} for a review).

The Anderson model of a localized magnetic impurity yields both the Kondo regime and the closely related Mixed Valence regime \cite{Haldane1978}. In the conventional Kondo regime the charge of an impurity or a nanostructure is an integer while the spin state alternates. In the Mixed Valence regime the charge is not quantized and also fluctuates \cite{Hewson}. In the Coulomb Blockade systems, the Mixed Valence regime has been previously realized in a narrow range of gate voltages close to the charge degeneracy points ({\it i.e.} in the vicinity of conductance peaks) \cite{GoldhaberPRL1998}. However, if the coupling to the contacts is increased, the regions of well-defined charge should gradually disappear and the Mixed Valence behavior should spread over the conductance valleys. Our paper is devoted to studying this regime.

We study the evolution \cite{note1} from the Kondo to the Mixed Valence regime by controlling the contact transparency within the same semiconducting  nanotube. The quantum-mechanical orbitals originating in two electronic subbands of nanotubes are doubly degenerate, forming four-electron ``shells'' (Figure 1) \cite{Liang2002,Buitelaar2002} (see also Ref. \cite{Alex2006} for additional references). The Kondo effect in this situation is expected to obey the SU(4) symmetry \cite{SU4Kondo,Choi2005}, as studied in Ref. \cite{Jarillo-Herrero2005} for one electron in a shell (see also Ref. \cite{Sasaki2004}). At low enough temperatures and sufficiently open contacts, the Kondo behavior develops in the valleys with one, two, and three electrons in the topmost shell \cite{Liang2002,Babic2004}. As the contacts are opened even more, so that the individual charge states are no longer well-defined, we observe the Mixed Valence behavior throughout the entire shell. In this regime, the four single-electron conduction peaks in a shell, visible at high temperature, merge at low temperatures into a single broad maximum. We successfully describe the conductance dependence on gate voltage by a simple formula (1), which we argue has deep implications. Finally, the characteristic temperature and energy scale in the Mixed Valence regime are found to be surprisingly small.  

\begin{figure}[h]
\includegraphics[width=0.80\columnwidth]{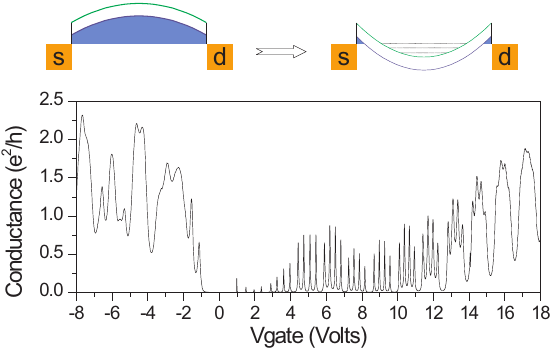}
\caption{\label{fig:ambipolar} Differential conductance of a 200 nm long small-gap semiconducting nanotube. Schematic: Band structure of a semiconducting nanotube, which can sustain either electron or hole populations, depending on the gate voltage. In the p-doped regime (negative $V_{gate}$) the nanotube conductance is relatively high. In the n-doped regime, the nanotube conduction is limited by the p-n junctions formed near the contacts. The resulting quantum dot conduction demonstrates a pronounced Coulomb blockade pattern ($V_{gate}>0$). As $V_{gate}$ grows, the tunneling barriers become narrower and more transparent, resulting in wider single-electron conduction peaks. }
\end{figure}

The nanotubes are grown on a Si/SiO$_2$ substrate by Chemical Vapor Deposition using CO as a feedstock gas \cite{Zheng2002}. Cr/Au electrodes are deposited on top of the nanotubes. We choose small-gap semiconducting nanotubes, which demonstrate high p-type conductance at negative gate voltages (Figure 1) \cite{Park2001,Cao2004}. At positive gate voltages, the middle section of the nanotube fills with electrons. The parts of the nanotube adjacent to the electrodes stay p-type (forming ``leads''). Thereby, a quantum dot is formed {\it within} the nanotube, defined by p-n and n-p junctions (schematics in Figure 1). As a result, Coulomb blockade sets in at low temperatures (Figure 1). (It is important for the observation of the SU(4) symmetry that the ``leads'' to the dot are formed within the same nanotube, and thus have the same orbital symmetry \cite{Choi2005}.) We choose to show the data measured on the same 200 nm - long nanotube in several cooldowns. Similar behavior was observed in a number of other samples. 

Due to the presence of the four-electron shells, the single-electron conduction peaks in Figure 1 ($V_{gate}>0$), cluster in groups of four. The peaks splitting within a shell is proportional to the charging energy of $E_C$, while the peak splitting between the neighboring shells is larger due to the additional shell splitting $\Delta$. The width of the p-n junctions in the nanotube depends on the gate voltage: larger positive gate voltages make the junctions narrower (see schematic in Figure 1). We therefore can vary the junctions transparency by changing the gate voltage. Indeed, the widths of the single-electron peaks in Figure 1 grows with the gate voltage, indicating the growing lifetime broadening of the levels, $\Gamma$. This extra control parameter allows us to study the evolution of transport regimes within the same sample. 

\begin{figure}
\includegraphics[width=3.3 in]{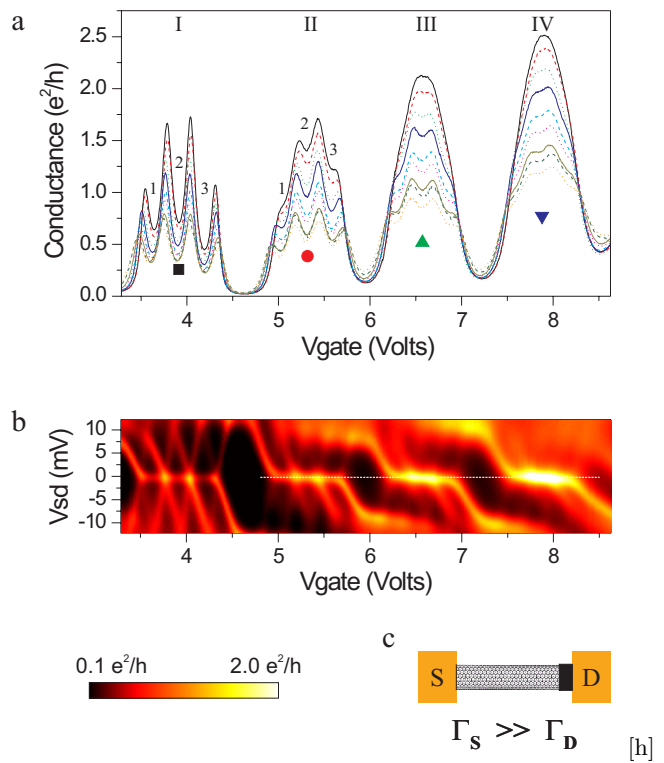}[h]
\caption{\label{fig:main} 
a) Differential conductance measured as a function of $V_{gate}$ at several temperatures (top to bottom: 1.3, 2.0, 3.3, 4.8, 6.4, 8.4, 10.4, 12.9 and 15.0 K). Four shells are shown (I-IV); the numbers (1-3) indicate the number of electrons in a shell. The Kondo effect enhances the conductance in the valleys within each shell at low temperatures. Eventually, the single-electron conductance peaks in shells III and IV merge to form smooth oscillations with a four-electron periodicity. 
b) Differential conductance map (same sample as in Fig. 2a)as a function of $V_{gate}$ and source-drain voltage (Scale of the colormap: 0.1 to $2e^2/h$; $T=3.3$ K). The outlines of the ``Coulomb diamonds'' \cite{QDreview} are visible, most clearly in shell I. The Kondo ridge is visible at zero source-drain voltage in all three diamonds. Going from shell I to shell IV, the Kondo ridge going across each shell becomes the most prominent feature of the data. (Figure 2a can be viewed as a horizontal cross-section of Figure 2b.) 
c) Schematic: In these data, one of the contacts is better coupled to the nanotube than the other, which results in up-down asymmetry of the image. One can consider the weakly coupled contact as a tunneling probe, which measures the density of states in the system made of the nanotube and the strongly coupled contact.}
\end{figure}

Figure 2a shows the temperature dependence of conductance in the gate voltage range covering four shells, marked I-IV. This Figure demonstrates a behavior commonly associated with the Kondo effect: the   conductance in the valleys grows at lower temperatures, resulting in the gradual disappearance of the single-electron peaks. In particular, the Kondo behavior is observed in the two-electron valleys, indicating a degenerate ground state. In Ref. \cite{Alex2006} we experimentally found that the energy splittings between the six possible states of two electrons in a shell \cite{Liang2002,Babic2004} are small. Hence even for a moderate broadening $\Gamma$, these states of two electrons are effectively degenerate and can all participate in the formation of the Kondo resonance \cite {Galpin2005,Izumida99}. Note, that this behavior is different from the one reported in Ref. \cite{Jarillo-Herrero2005}, where the six states of two electrons are non-degenerate.

The lifetime broadening $\Gamma$ ranges from $\sim 7$ meV in shell I to $\sim 15$ meV in shell IV. (We extract this information by fitting multiple Lorentzian peaks to the highest temperature curve. Within each shell, the peaks have similar $\Gamma$.) This broadening should be compared to other relevant energy parameters: the charging energy of $E_C \approx 10$ meV and the shell spacing of $\Delta \approx 10$ meV that we extract from Figure 2b. Therefore the charge fluctuations should be significant, in particular in shells III and IV, where $\Gamma > E_C$. Thus, the Mixed Valence regime should cover the entire $V_{gate}$ range in these shells \cite{note1}.

Experimentally, the growth of conductance in the valleys completely washes away the single-electron peaks at the lower temperatures in shells III and IV (Figure 2a)  However, the unitary limit plateaus, observed for a well-developed Kondo effect \cite{vanderWiel2000}, are absent. Instead, we find deep oscillations with a periodicity of four electrons and the maximal conductance exceeding $2e^2/h$.
From the nonlinear transport measurement (see the discussion of Figure 2b below), we can estimate \cite{Babic2004} the ratio of the coupling to the source and the drain $\Gamma_S / \Gamma_D \sim 4$ for shell IV. We expect the maximal conductance (for 4 modes) to be given by $4G_0=4e^2/h \frac{4\Gamma_S \Gamma_D }{(\Gamma_S+\Gamma_D)^2} \approx 2.6 e^2/h$, in good agreement with experiment.

The map of conductance as a function of the source-drain voltage $V_{SD}$ and gate voltage $V_{gate}$ (same range as in Figure 2a) is shown in Figure 2b. In the first four-electron shell, the ``Coulomb diamonds'' are clearly visible. A ridge at zero source-drain voltage, typical for the Kondo effect \cite{Kondo1998}, forms across the three inner valleys of shell I \cite{Liang2002,Babic2004}, and further grows in shells II-IV to become the most prominent feature of shells III and IV. 

In this sample, one of the contacts is more strongly coupled to the nanotube than the other, as evident from the lack of up-down symmetry of the signal strength in Figure 2b. The weakly coupled contact serves as a tunneling probe that senses the density of states in the nanotube strongly hybridized with the Fermi-sea of the other contact \cite{Pustilnik&Glazman2004}. Therefore, the conductance dependence on $V_{SD}$ may be viewed as a qualitative measure of the density of states in the SU(4) Kondo or Mixed Valence regimes. The conductance is enhanced whenever the Fermi level of the weakly coupled contact is aligned with a resonance in the density of states of the nanotube. In shells III and IV, a single resonance descends from the positive source-drain voltages to the Fermi energy as the gate voltage increases. The resonance dwells there (forming a Kondo ridge), and then continues the descent to negative voltages. We observe that the Kondo resonances for 1 and 3 electrons are shifted respectively above and below zero bias. As a result, the Kondo conductance ridges acquire a small slope, noticeable in shells II, III and IV of Figure 2b.

The smooth conductance modulation with a period of four electrons, as observed at lowest temperatures in Figure 2a, is superficially similar to the single-particle (Fabry-Perot) interference \cite{Liang2001}. Nevertheless, there are striking differences between the two regimes. First, the conductance map as a function of the source-drain and gate voltages in the single-particle regime demonstrates a crisscrossing grid of straight lines, corresponding to the (anti-) resonance interference conditions \cite{Liang2001}. In our case, the conductance map is distinctly different: the positions ($V_{SD}$ vs. $V_{gate}$) of the conductance resonances depend non-linearly on the gate voltage (as discussed above; Figure 2b), indicating the importance of electronic interactions. Second, the temperature evolution of conductance in Figure 2a, indicates the presence of a new energy scale ($\sim $1 meV), which is significantly smaller than the single-particle energies (the shell spacing $\Delta$ and the charging energy $E_C$ are both approximately equal to 10 meV). Third, the Fabry-Perot behavior is established for conductance oscillations, which have a small amplitude ($\sim 10 \%$ of the background). We observe deep conductance variations between the centers of the four-electron shells ($\sim 2e^2/h$) and the valleys between the shells ($\sim 0.2e^2/h$). 

\begin{figure}
\includegraphics[width=1.0\columnwidth]{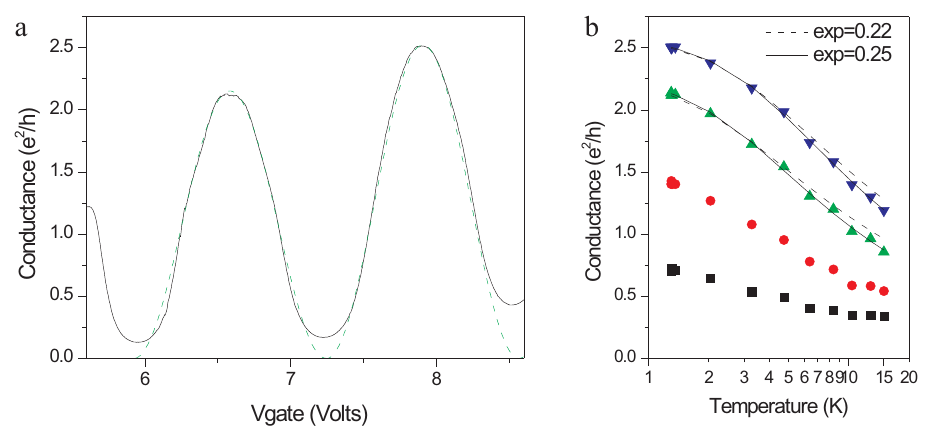}
\caption{\label{fig:temper} 
a) Conductance in shells III and IV at $T=1.3$ K (solid black line) compared to the results of formula (1) (green dashed line). The prefactor ($4G_0$) is adjusted for shells III and IV separately to match the corresponding conductance maxima. The $V_{gate}$ scale is uniquely fixed by the 4-electron periodicity of the features. 
b). Symbols: temperature dependence of the conductance in the two-electron valleys (shells I-IV, bottom to top; the centers of the shells are marked by the corresponding symbols in Figure 2a). Lines: fit using the empirical function of Ref. \cite{GoldhaberPRL1998} (dashed blue line) and similar fit with an exponent of 0.25 instead of 0.22 (solid black line).}
\end{figure}

We have successfully fitted the lowest temperature conductance for shells III and IV (Figure 3a) by a remarkably simple formula:  
\begin{eqnarray} 
&&
G \sim \sin^2 [\pi C_{gate} (V_{gate}-V_{gate}^{(0)})/4].
\label{eq:E1}
\end{eqnarray}
Here $N = C_{gate} (V_{gate}-V_{gate}^{(0)})$ is the number of electrons on the topmost shell, which is assumed to change continuously with gate voltage \cite{GlazmanHekkingLarkin1999}. Although this functional form may seem deceptively simple, its implications are far reaching: Kubo's formula predicts conductance in the form of $G = G_0 \sum_i \sin^2 (\delta_i )$, where $\delta_i $ are the scattering phase shifts of different electron modes (enumerated by index $i$), traversing the nanostructure \cite{Pustilnik&Glazman2004}. In case of a nanotube, there are four modes ($i=$1-4), corresponding to two spin projections and two subbands. Friedel's sum rule \cite{Pustilnik&Glazman2004}, yields a relation between the number of electrons added to a nanostructure and the sum of the scattering phase shifts: $\sum_i \delta_i = \pi N$. If we assume that the scattering phases for all modes are equal to each other, we obtain $\delta_i =\pi N /4$. The conductance then becomes $G =4 G_0 \sin^2 (\pi N/4)$, in accord with our observations. 

We stress that $\delta_i =\pi N /4$ is not obvious {\it a priori}. In the Kondo regime (integer $N$), the scattering phase shifts of different modes are expected to become equal to each other \cite{Pustilnik&Glazman2004}. However, in our case, this result holds also for {\it non-integer} $N$'s. We would also like to point out that formula (1) represents an {\it essentially} many-body behavior. Indeed, it successfully describes the conductance only at the lowest temperatures ($T \lesssim 5$ K), much smaller than $\Gamma$, $E_C$ and $\Delta$. Already at slightly higher temperatures, the single-electron features appear in the $G(V_{gate})$ curve, indicating that formula (1) breaks down.

In Figure 3b we plot the conductance measured in the middle of the two-electron valleys (locations marked by symbols on Figure 2a) as a function of temperature. The high-temperature conductance for shells III and IV and the low-temperature conductance for shells I and II shows the logarithmic dependence expected for Kondo effect \cite{note1}. We have fitted the temperature dependence of conductance for shells III and IV using the empirical formula of \cite{GoldhaberPRL1998}, used to fit the Numerical Renormalization Group data of \cite{Costi1994}. We find that the observed dependence is slightly steeper then dictated by the original formula with an exponent of 0.22 and is better described by an exponent of 0.25 for both shells. (The difference is not large, but visible.) 

The half-width of the Kondo ridge (2 meV) measured across the center of shell IV (Figure 2b, $V_{gate}$ = 7.8 V) agrees well with the characteristic temperature ($T_0 = 20$ K) extracted in Figure 3b. (One can estimate $T_0$ from Figure 3b as the temperature by which the conductance drops by one half.) The relation between $T_0$ and other energy parameters is not precisely known in the SU(4) Mixed Valence regime; however, it seems surprising that the measured $T_0$ is $\sim 5$ times smaller than $\Gamma$ (which is about the same as both $\Delta$ and $E_C$). In our view, understanding the emergence of the small energy scale $T_0$ in the Mixed Valence regime remains a theoretical challenge.

In conclusion, we identify a transport regime, where the low temperature conductance shows no single electron features, which counterintuitively appear at high temperatures. This Mixed Valence regime differs from the related Kondo regime by a stronger electron tunneling to the contacts, which smears the charge quantization in the quantum dot. We successfully account for the low-temperature conductance as a function of gate voltage in the new regime. We find that the energy scale $T_0$, which controls the temperature evolution of conductance and the tunneling density of states is about an order of magnitude smaller than expected. Studies of the Mixed Valence regime in nanostructures may offer additional clues to understanding strongly correlated electronic systems, such as heavy fermions and mixed valence compounds. 

Recently, we have investigated the SU(4) Kondo behavior for one, two and three electrons in magnetic field \cite{SU4PRB}. We have obesrved the transitions from the SU(4) to both the spin and the orbital SU(2) Kondo effects.

Acknowledgements: We thank H. Baranger, A. Chang, L. Glazman, E. Novais, K. Matveev, G. Martins, M. Pustilnik, and D. Ullmo for valuable discussions. The work is supported by NSF DMR-0239748.

    \begin{widetext}
    
    \begin{figure}[ht]
    \includegraphics[width=1\columnwidth]{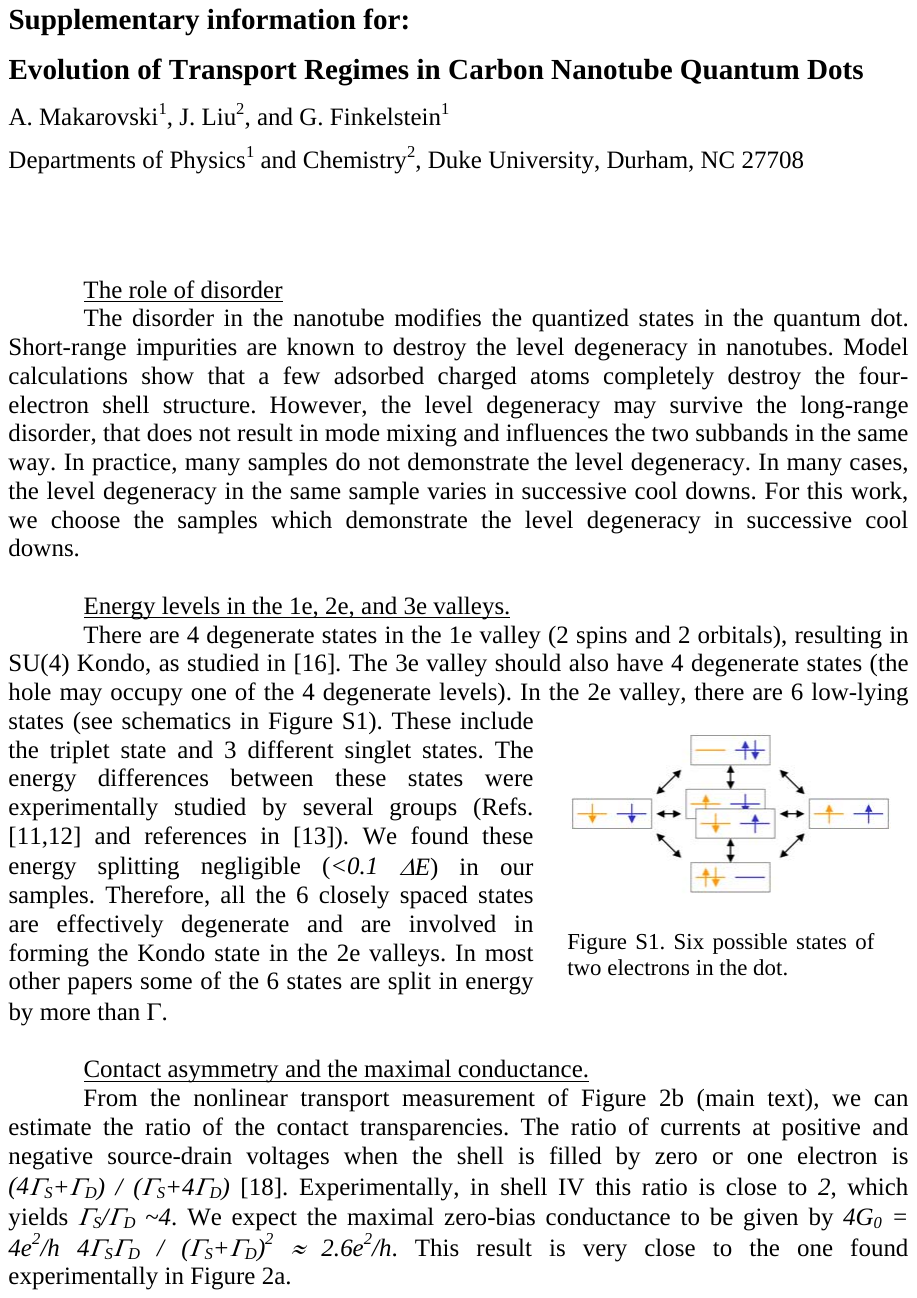}
    \end{figure}
      
\end{widetext}  

\begin{widetext}

     \begin{figure}[ht]
     \includegraphics[width=1\columnwidth]{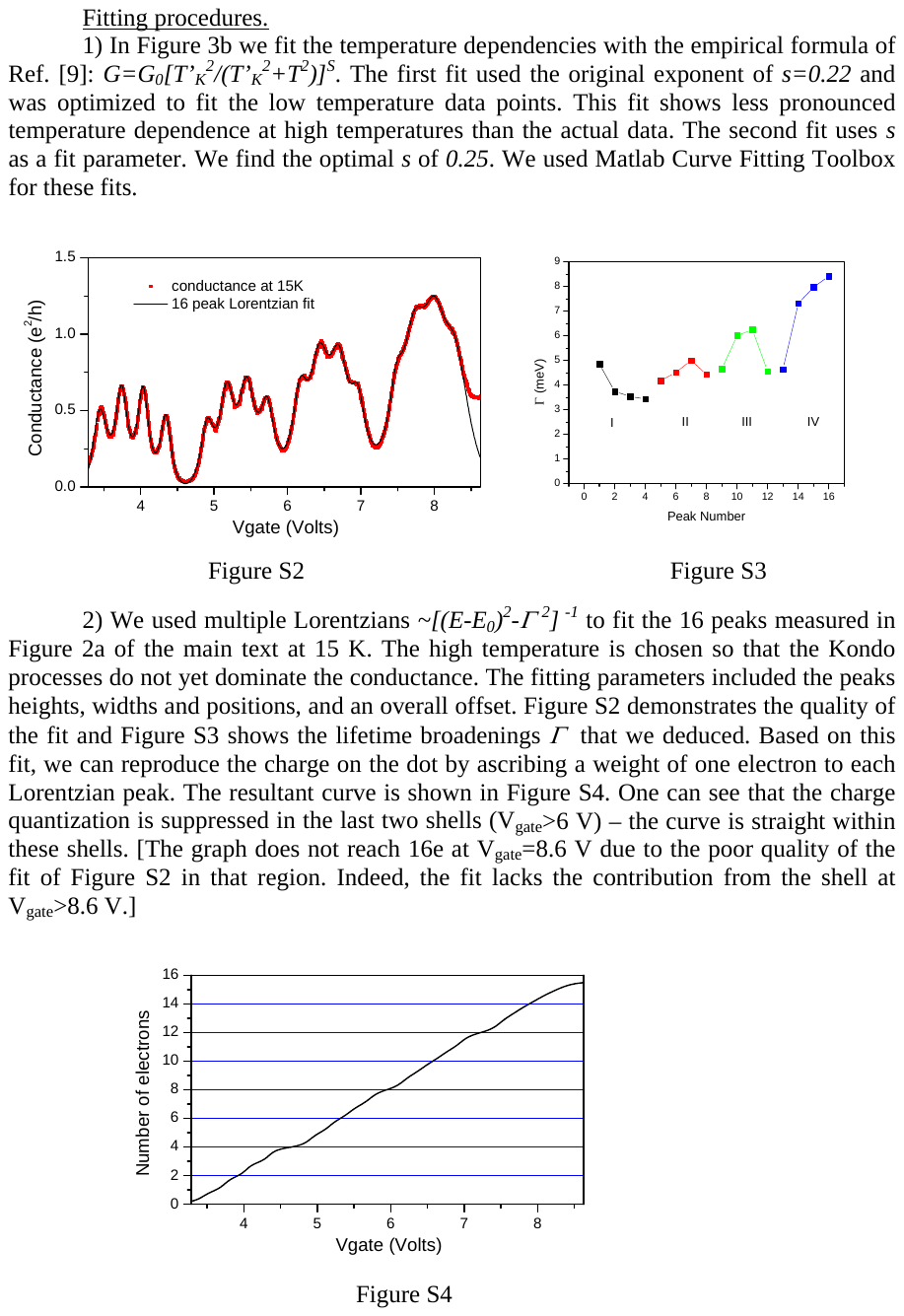}
     \end{figure}
     
 \end{widetext}

\end{document}